\documentstyle[preprint,aps]{revtex}
\begin{document}
\draft
\title{Theta-point universality of random polyampholytes
with screened interactions}
\author{Pietro Monari$^1$ and Attilio L. Stella $^{1,2}$}
\address{$^1$INFM -- Dipartimento di Fisica,
Universita' di Padova, I--35131 Padova, Italy}
\address{$^2$ Sezione INFN, Universita' di Padova, I--35131 Padova, Italy}
\maketitle
\begin{abstract}
By an efficient algorithm we evaluate exactly the disorder-averaged 
statistics of globally neutral self-avoiding chains with quenched 
random charge $q_i=\pm 1$ in monomer $i$ and nearest neighbor 
interactions $\propto q_i q_j$ on square (22 monomers) and cubic 
(16 monomers) lattices.  At the theta transition in $2D$, radius 
of gyration, entropic and crossover exponents are well compatible 
with the universality class of the corresponding transition of 
homopolymers. Further strong indication of such class comes from 
direct comparison with the corresponding annealed problem. 
In $3D$ classical exponents are recovered.  The percentage of charge 
sequences leading to folding in a unique ground state approaches zero 
exponentially with the chain length.
\end{abstract}
\vspace{1cm}
\narrowtext

During the last years random heteropolymers have been 
the object of intense study: they play a central role
for our understanding of the properties of biologically 
active molecules \cite{cariche} and represent a relatively 
handable model of  disordered system, of great interest 
in statistical mechanics \cite{glass,frau,garel}.
Particular attention has been focused recently on randomly 
charged polymers (polyampholytes)\cite{gen} interacting via 
long--range (Coulomb) and short--range (screened) potentials. 
Aminoacids in proteins carry electric charges and electrostatic  
interactions play an important role in determining their behavior 
\cite{cariche}.  Thus, polyampholyte models are expected
to be useful for the investigation of biologically relevant 
polymers.

Due to monomer--monomer and monomer--solvent interactions, even 
in absence of charges, a linear polymer undergoes a collapse 
$\Theta$-transition as the temperature $T$ is varied \cite{carlo}. 
In particular, in the case of a homopolymer, for $T>T_{\Theta}$ 
the chain is swollen and its average radius of gyration grows as 
$R_g \propto N^{\nu}$ ($N$ is the number of steps, $\nu=3/4$ in $2D$ 
\cite{ninh} and $\nu=0.588$ in $3D$\cite{zinn}), while for $T<T_{\Theta}$ 
the polymer is compact ($\nu=1/d$).  At $T=T_{\Theta}$ a distinct 
universality class ($\nu=\nu_{\Theta}=4/7$ in $2D$ \cite{carlo} 
and $1/2$ in $3D$ \cite{sal}) characterizes scaling.

The existence of a collapse transition is by now well established 
for polyampholyte models with both short-range and long-range 
interactions due to randomly distributed 
charges. In the long-range Coulomb case,
the total charge distributed along the chain must not exceed a critical 
value $( \Delta Q \simeq \sqrt{N})$ in order for the $\Theta$-transition 
to occur \cite{kk2}.  In the short-range case there is a collapse  transition
as long as the charge unbalance $x=\vert N_{+}-N_{-} \vert/(N_{+}+N_{-})$
is less than some cutoff value \cite{kk1}.  The collapse of randomly charged 
polymers owes much of its importance to the close connection with protein 
folding\cite{cre2} and is similar to that of homopolymers, 
with a $\Theta$-scaling regime separating compact and swollen phases.

An important, yet unsettled issue is that of establishing whether the
$\Theta$-transition of polyampholytes falls in the same universality class
as the collapse of homopolymers described above. 

By a numerical study of
randomly charged self-avoiding walks (SAW) with nearest neighbor 
interactions on the square lattice, Kantor and Golding \cite{kg} got
$\nu_{\Theta}=0.60\pm 0.02$ for  globally neutral polyampholytes, slightly, 
but appreciably different from that of homopolymers ($\nu_{\Theta}=4/7$).
These results, based mostly on MC enumerations (up to $N=99$) and on a 
relatively limited sampling over disorder, suggest that the presence 
of quenched random interactions can modify the universality class of 
the $\Theta$-transition.  This is a quite intriguing possibility, also 
in view of the constant focus on universality issues in the literature 
on $\Theta$-transitions \cite{carlo}.  A major limitation of this kind 
of studies is due to the need of performing averages over charge disorder, 
a task which, at an exact level, becomes impossible, with standard 
algorithms, as soon as $N>15$ in $2D$. It is not obvious whether, 
for relatively short chains, quenched averages over few disorder configurations 
($10 \div 100$ in ref\cite{kg}) should be  sufficient for a satisfactory 
determination of the radius of gyration.  Especially in the low--temperature 
regime, fluctuations of thermodynamic quantities due to quenched disorder are 
indeed very large and random sampling over a small number of disorder 
realizations can lead to inaccurate results for quenched averages.  
The investigation of the low--temperature phase of random polyampholytes 
is a formidable challenge for MC methods, because of the prohibitively 
large samplings it should require.  On the other hand, even thermal averages 
become a problem at low T. Indeed, dynamic MC methods like the pivot algorithm  
are efficient in the high temperature regime but much less reliable as 
soon as the temperature is lowered below $T_{\Theta}$ and chains are 
compact\cite{whit}. Finally, Markov chain based MC methods do not allow
computation of entropic exponents, which would provide a more complete 
characterization of the universality of the transition.

The above considerations suggest that an interesting strategy in the study
of random polyampholyte models can be that of extending as far as
possible, by appropriate algorithms, the range of chain lengths within which
we can still gather exact information.  In the present work we perform exact 
enumerations up to chains with $N=21$ in $2D$ and $N=15$ in $3D$
for polyampholytes with charge disorder and nearest neighbor interactions
and carry out quenched averages over all disorder realizations.
To this purpose we developed a powerful algorithm for short-range interacting
SAW's with charge disorder, in such a way as to reduce by orders of magnitude 
 the required computational effort, compared to more standard methods.

We represent each  $N$-steps  polymer chain configuration by a SAW $\omega$ 
$(\vert \omega \vert=N)$ on square or cubic lattice. In each of the 
$N+1$ visited sites sits a charged monomer.  The hamiltonian takes the form:

\begin{equation}
H(\{q\},\omega)=-\sum_{\langle i,j\rangle} q_{i}q_{j}, 
\label{ham}
\end{equation}

\noindent{where $q_i=\pm1$ is the charge carried by the i-th monomer and
$\langle i,j \rangle$ indicates pairs of (non-consecutive) nearest 
neighbor monomers. The sequence of the $N+1$ charges distributed 
along the chain is denoted by $\{ q \}=\{q_{0},q_{1},...,q_{N}\}$          
and is assumed to be globally neutral ($N+1$ even).  The quenched 
average squared radius of gyration is:}

\begin{equation}
R_{g}^2(N,T)=
\frac{1}{N_{q}}
\sum_{\{q\}}
Z_{\{q\}}(N,T)^{-1}
\left [
{\sum_{\vert \omega \vert=N}e^{-H(\{ q \},\omega)/T}
r_{g}^2(\omega)}
\right ],
\label{raggio}
\end{equation}

\noindent{where $ N_q$
is the total number of charge sequences, $r_g(\omega)$ is the radius 
of gyration with respect to the center of mass of the configuration 
$\omega$, and $Z_{\{q\}}(N,T)$ is the canonical partition function
for a given realization $\{q \}$ of disorder:}

\begin {equation}
Z_{\{q\}}(N,T)=\sum_{\vert \omega \vert=N}e^{-H(\{q\},\omega)/T}. 
\label{zetaqu}
\end{equation}

Near the $\Theta$-point the radius of gyration is expected to obey the 
tricritical scaling form \cite{janni,dege}:

\begin {equation}
R_{g}^2(N,T) \simeq N^{2\nu_{\Theta}} f(N^{\phi_{\Theta}}\tau),
\label{tric}
\end {equation}

\noindent where $\tau = (T-T_{\Theta})/T_{\Theta}$ is the temperature 
distance from the $\Theta$-transition and  $\phi_{\Theta}$ is the
 crossover exponent.
In the case of homopolymers in $2D$, $\nu_{\Theta}=4/7$ and $\phi_{\Theta}
=3/7$ \cite{sal}.

The annealed partition function is defined as the average
of (\ref{zetaqu}) over all sequences:

\begin{equation}
 Z_{(a)}(N,T)= \frac{1}{N_{q}}\sum_{\{q\}} Z_{\{q\}}(N,T) 
\label{partann}
\end{equation}

\noindent while the  quenched partition function is defined 
in terms of the quenched free energy: 

\begin{equation}
 Z_{(q)}(N,T)=
\exp{ \left [ \frac{1}{N_{q}}\sum_{\{q\}} \log{(Z_{\{q\}}(N,T))} 
\right ]}.
\label{partque}
\end{equation}

Both annealed and quenched partition functions are expected to scale as:

\begin{equation}
Z_{(a/q)}(N,T)\simeq \mu_{(a/q)}(T)^{N} N^{\gamma_{(a/q)}(T)-1}. 
\label{scalz}
\end{equation}

The connectivity $\mu$ is  lattice and temperature dependent,
while $\gamma$ is the  universal entropic exponent.  Like in the 
homopolymer case, for high T the exponent $\gamma(T)$ should identically 
take the value appropriate to SAW's in the swollen regime $(\gamma_{SAW}
=43/32)$\cite{ninh}.  In presence of a $\Theta$-collapse, at $T_{\Theta}$, 
$\gamma$ is expected to assume a different value, $\gamma_{\Theta}$, 
which is peculiar of the universality class of the transition.
At lower temperatures, because of the globular shapes of the collapsed 
polymer, surface effects are present and can modify the above scaling form 
for $Z$ with the appearance of an extra exponential factor, besides 
$\mu^N$ \cite{freez}.

It is not obvious, a priori, that $\mu$ and $\gamma$ should take the same 
values for the quenched and annealed problems.  While this is plausible at 
relatively high T, where quenched disorder plays a minor role, discrepancies 
can be anticipated at low T. A main issue here is to establish whether 
$T_{\Theta}$ is still included in the high-T range.

Different entropic exponents have to be defined when polymers are subject  
to geometrical constraints: if the polymer chain  is forced to live
in half space by an impenetrable wall to which one of its ends is fixed, the
critical entropic exponent assumes a value $\gamma_1$, different from the 
bulk $\gamma$ \cite{carlo}.  
The entropic behavior of SAW's at boundaries already played a major role
in studies aimed at a precise characterization of the universality
class of the $\Theta$--transition of homopolymers (in $2D$:  
$\gamma_{\Theta}=8/7$, $\gamma_{1\Theta}=4/7$, $\gamma_{11\Theta}=-4/7$
\cite{vander,flavio}).

The numerical study of entropic exponents is greatly facilitated by 
considering simultaneously data for bulk and boundary behavior\cite{euro}.
Indeed, if the polymer is not adsorbed, the connectivity $\mu$ is 
insensitive to the presence of boundary and remains the same for both
bulk and boundary behavior of $Z$.  Below we indicate by $Z^{bulk}$ and 
$Z^{half}$ the respective partition functions.  Thus, the ratio between
bulk and boundary $Z$'s  scales as a power of the difference between 
the respective $\gamma$'s and does not depend on $\mu$, whose estimation 
is then not necessary.  Due to these circumstances, the determination of 
$\gamma-\gamma_1$ gets easier and much more accurate.

Here we call contact a pair of non-consecutive monomers on nearest-neighbor 
sites, i.e. two interacting monomers.  The contact map of a given SAW 
configuration $\omega$  is the  set of all contacts it contains:

\begin{equation}
X(\omega)=\left \{(i,j):\vert \omega(i)-\omega(j)\vert =1,\vert i -j \vert
> 1 \right \}. 
\label{set}
\end{equation}

A   contact map of an $N$-steps SAW can also be  represented by an 
$(N+1)\times (N+1)$ matrix, whose $(i,j)$ element is 1 or 0, according 
to whether the monomers $i$ and $j$ are interacting or not , respectively.
For any given $\{q\}$, the energy of a configuration $\omega$ is fully 
determined by its contact map $X(\omega)$.  Two configurations $\omega$, 
$\omega '$ which are characterized by the same structure of contacts 
($X(\omega)=X(\omega ')$), have the same energy for every $\{q\}$,
and can be considered  as equivalent.  The set of all $\omega$'s of 
a given length can be partitioned into equivalence classes, each of 
them containing all the  walks which are characterized by a given 
contact map. The number of equivalence classes is equal to the number, 
$S_{N}$, of distinguishable contact maps $X_{\alpha},~\alpha=1,..,S_N$.
Each equivalence class $C_{\alpha}=\left\{\omega:X(\omega)=X_{\alpha}\right\}$
is characterized by its own degeneracy $g(\alpha)$ and cumulative
squared radius of gyration $\rho_g^2(\alpha)$:

\begin{equation}
g({\alpha})=\sum_{\omega \in C_{\alpha}}1;
~~~~
{\rho}_{g}^2({\alpha})=\sum_{\omega \in C_{\alpha}}r_g^2(\omega).
\label{dege}
\end{equation}

\noindent
$ g(\alpha)$ is expected to grow exponentially with the difference 
between the number of steps $N$ and the number of contacts in
$ X_{\alpha}$ \cite{ishi}.  This means that $S_N$ still grows 
exponentialy with $N$, but much more slowly than the total number 
of SAW's $C_N$ (see Table 1).  In particular the ratio $S_N/C_N$ 
is expected to approach zero exponentially.  

The sum over configurations
$\omega$ in (\ref{raggio}) and (\ref{zetaqu}) can be replaced by the 
sum over equivalence classes, each of them taken with its own
degeneracy and cumulative squared radius of gyration:

\begin{equation}
R_{g}^2(N,T)=
\frac{1}{N_{q}}
\sum_{\{q\}}
Z_{\{q\}}(N,T)^{-1}
\left [
{\sum_{\alpha=1}^{S_N}e^{-H(\{ q \},X_{\alpha})/T}
{\rho}_{g}^2{(\alpha)}}
\right ];
\label{newraggio}
\end{equation}

\begin {equation}
Z_{\{q\}}(N,T)=\sum_{\alpha=1}^{S_N}e^{-H(\{q\},X_{\alpha})/T}g(\alpha). 
\label{newzeta}
\end{equation}

In terms of computational cost, the last equations are considerably 
cheaper
than eqs. (\ref{raggio}) and (\ref{zetaqu}).  The main improvement 
regards the thermal averages over configurations $\omega$, which are 
made considerably faster, due to the fact that they involve summations 
over $S_N$ rather than $C_N$ terms.  Detailed enumeration of all
$\omega$'s for each given sequence $\{q\}$, would become unfeasible
as soon as $N > 15$, when computing exact averages over disorder.  

In the present work, eqs.(\ref{newraggio}) and (\ref{newzeta}) have been 
implemented by an efficient algorithm, in which SAW's of a given length 
are generated once for all.  The structure of contacts of each walk  is 
registered on a binary map.  Whenever a new walk is generated,
its contact map is analyzed and sorted: if the contact configuration has 
already occured, its degeneracy and cumulative gyration radius 
are updated, otherwise a new contact map is added.

Once SAW's are fully enumerated, all contact maps $X_\alpha$ are stored 
together with their $g(\alpha)$ and 
$\rho_g^2(\alpha)$.  Disorder-averages of thermodynamic 
and geometric observables  are then calculated over half the number of 
neutral sequences, being the hamiltonian invariant under $\{q\}\to \{-q\}$.

On a DEC 600 DIGITAL workstation exact enumeration of SAW's and 
complete quenching over all sequences require few minutes of CPU time 
for 16 monomer chains, about 130 hours for 22 monomer chains.

The same algorithm was later on adapted in order to compute
annealed averages over the same realizations of disorder.  
The computation of annealed averages is slightly faster than 
that of quenched averages. Thus, we could easily obtain exact 
results with annealed disorder for $N$ up to 21 in $2D$.

In order to study the $\Theta$-point we computed effective $\nu$ exponents:

\begin{equation}
\nu(N,K,T)=\frac{1}{2}
\log{\left [ \frac
{R_{g}^2(N,T)}{R_{g}^2(N-K,T)}\right]}
{\log{\left [ \frac{N}{N-K}\right ]}}^{-1}.
\label{nupriv}
\end{equation}

In the $N\rightarrow \infty $ limit these curves should be step 
functions of T. However,  at finite $N$, they show a rounded step.  
If the trends of approach of the $N \to \infty$ $\nu$ values in the 
high T and low T phases are from opposite directions, the curves 
(\ref{nupriv}) are expected to intersect among themselves in the 
neighborhood of the $\Theta$-point.  They  indeed show such behavior: 
effective exponents like $\nu(N,2,T)$ are monotonically increasing 
functions of $N$ at high T and decreasing at not too low T. Linear 
extrapolation of these curves with respect to $1/N$, in the $1/N \to 0$ 
limit, allows to estimate an exponent $\nu_{\infty}(T)$ which is close 
to or even 
below the compact-polymer value $\nu=0.5$ for T just below the 
intersection region.  On the other hand, $\nu_{\infty}(T)$  is almost 
equal to the swollen SAW value $\nu=0.75$ at high T (Fig.1).   Intersections 
of all the curves $\nu(N,K,T)$ occur in a small region of the 
$(T, \nu)$ plane, within which one can suppose the transition 
to be located.  Following Privman \cite{priv}, in order to
quantitatively pinpoint the $\Theta$-transition, we calculated 
the coordinates $(T_{int},\nu_{int})$ of all intersections
between every pair of curves $\nu(N,K,T)$ , $\nu(N{\rm '},K{\rm '},T)$,
and plotted these points against $1/N_{eff}=2/(N+N{\rm '})$.
The definition of $N_{eff}$ is of course subjective.  In our 
choice  no role is played by the integers $K$ and $K{\rm'}$
because of the weak dependence of the intersection locations
on these parameters.  For each $1/N_{eff}$ we computed the mean 
of $T_{int}$ and $\nu_{int}$ of the corresponding intersections 
and  extrapolated them linearly as a function of $1/N_{eff}$ 
(Fig.2) obtaining  the estimates  : $\nu_{\Theta}= 0.58 \pm 0.02$ 
and $T_{\Theta}=0.80\pm 0.03$.  Uncertainty estimates are also based 
on comparison between extrapolations from data in different ranges 
of $1/N_{eff}$.  The exponent is fully compatible with homopolymer 
$\Theta$-point universality.

Another method can be applied in order to estimate $\nu_{\Theta}$ and
$T_{\Theta}$. As illustrated above, the effective exponents 
$\nu_N(T)=\nu(N,2,T)$ 
are monotonic functions of $1/N$, decreasing for $T>T_{\Theta}$ and
increasing  for $T<T_{\Theta}$ .
Their linear correlation 
with respect of $1/N$ can be analyzed with the correlation coefficient
defined by \cite{feller}:
\begin{equation}
r(T)=\frac
{\sum_N (1/N-\overline{1/N})(\nu_N(T)-\overline{\nu_N(T)})}
{\sqrt{{\sum_N (1/N-\overline{1/N})}^2
{\sum_N (\nu_N(T)-\overline{\nu_N(T)})}^2}},
\label{rcorr}
\end{equation}
where bars indicate averages over $N$.
The coefficient $r(T)$ is close to $-1$ for $T>T_{\Theta}$ 
and to $1$ for $T<T_{\Theta}$ meaning that,
in these regions,  data are very well linear correlated 
and have opposite monotony. 
In the $\Theta$ region  $r(T)$ undergoes a sudden jump between 
$1$ and $-1$.
Its derivative with respect to temperature shows a 
high and sharp peak whose mean value and width localize $T_{\Theta}$ and 
determine its uncertainty $\Delta T_{\Theta}$.  The extremal values taken 
by $\nu_{\infty}(T)$ in the interval $[T_{\Theta} -\Delta T_{\Theta},T_{\Theta}
+ \Delta T_{\Theta}]$ give an estimate of $\nu_{\Theta}$ and of the 
corresponding error $\Delta \nu_{\Theta}$.  $T_{\Theta}$ and $\nu_{\Theta}$ 
obtained with this method are almost identical to the values determined above 
by extrapolating the intersections $\nu_{int}$ and $T_{int}$, respectively.

In order to obtain the crossover exponent $\phi_{\Theta}$ we analyzed 
the derivative of the squared radius of gyration with respect to 
temperature. Near the $\Theta$-point, this quantity should scale as:

\begin {equation}
\frac{d}{dT} R_g^2(N,T) \simeq N^{\phi_{\Theta}(T)+2{\nu(T)}}.
\label{corr}
\end {equation}

\noindent
 
The effective exponent curves corresponding to $\phi + 2 \nu$ 
do not clearly intersect each other in a narrow region of the $(T,\phi+2\nu)$
plane. So, the method used for determining $\nu_{\Theta}$ cannot be applied 
in this case, 
because it would lead to ambiguos results.  Following ref. \cite{priv}, we 
then calculated, for each intersection $(T_{int},\nu_{int})$ between 
$\nu(N,K,T)$ and $\nu(N{\rm '},K{\rm '},T)$, with $N>N{\rm '}$, the quantity:

\begin {equation}
\log{\left[\frac{dR_g^2(N,T_{int})/dT}{dR_g^2(N-K,T_{int})/dT}\right]} 
{\log{\left[\frac{N}{N-K}\right]}}^{-1}
-2 \nu_{int}.
\label{fipriv}
\end{equation}

Extrapolation of these data in $1/N_{eff}$ leads to the estimate: 
$\phi_{\Theta}=0.40\pm 0.08$ (Fig.3).  The computation   of the 
crossover exponent for homopolymer $\Theta$-transitions  
is usually rather difficult and often leads to 
considerable overestimates \cite{chang,grassb}. Our result is well 
compatible with the exact homopolymer value $\phi_{\Theta}=3/7\simeq 0.42..
$\cite{sal}.  Attempts to determine $\phi$ on the basis of data 
collapse fits for $R_g$ (eq.(\ref{tric})) were not very successful 
because the collapse quality does not depend sensibly enough on this exponent.

In the annealed system, frustration effects peculiar of quenched disorder 
are ruled out.  The charges distributed along the chain are indeed free 
to rearrange among themselves in such a way to let nearest-neighbor 
interactions able to minimize the energy of each SAW configuration 
$\omega$.  It seems very plausible that such a rearrangement can produce
a collapse in the same universality class as the $\Theta$-point of an
ordered polymer with nearest neighbor attractive interactions for
all monomers. Because of these reasons we expect annealed disorder to
be irrelevant for the collapse transition.  This  conjecture is well 
confirmed by  the analysis of our exact enumeration results for the annealed 
system (22 monomers).  The analysis followed the lines of the quenched case.
The transition exponents of the annealed model were estimated as 
$\nu_{\Theta}=0.58\pm 0.02$ and $\phi_{\Theta}=0.41 \pm 0.08$.

A direct comparison between annealed and quenched entropies turns then 
out to be a very significant test, in view of the fact that the 
annealed system represent a sort of substitute of the pure one. As 
explained above, to avoid difficulties due to the calculation of 
the non-universal constant $\mu$, in the case of both quenched and 
annealed charges, we analyzed the ratio $\mathcal{Z}$ between the partitions of 
SAW's in the bulk and in presence of boundary, which is expected to 
scale as:

\begin{equation}
{\mathcal{Z}}_{(a/q)}(N,T)
=\frac
{Z^{bulk}_{(a/q)}(N,T)}{Z^{half}_{(a/q)}(N,T)}
\simeq N^{\left ( \gamma-\gamma_{1}\right )_{(a/q)}(T)}.
\label{zeta}
\end{equation}

Effective exponents can be obtained from:

\begin{equation}
\left [ \frac{{\mathcal{Z}}_{(a/q)}(N,T)}
{{\mathcal{Z}}_{(a/q)}(N-2,T)}\right ] ^{{1}\over{2}} 
= 1+\frac{1}{N}(\gamma-\gamma_1)_{(a/q)}(N,T)+
{\rm o}\left({\frac{1}{N^2}}\right).
\label{raf}
\end{equation}

The sequences $(\gamma-\gamma_1)_{(a/q)}(N,T)$,  plotted against $1/N$,
show remarkably good linear correlation. 
Their  extrapolation for $1/N \to 0$
gives a reasonable estimate of the expected $(\gamma-\gamma_1)_{(a/q)}$
in the high-T range and close to $T_{\Theta}$.
Even more precise is the comparison between the
annealed and quenched cases based  on these $\gamma - \gamma_1$
estimates.
It turns out  that the difference $\gamma -\gamma_{1}$
is almost identical for annealed and quenched systems on a range
of temperatures which clearly extends below the $\Theta$-temperature (Fig.4).
We estimated $(\gamma - \gamma_1)\sim 0.50$ and 
$(\gamma - \gamma_1)\sim 0.39$ at the $\Theta$-point and in the high T
region, respectively.  
The $\Theta$-point determination is slightly below the
homopolymer value 
($(\gamma-\gamma_1)_{\Theta}=4/7$\cite{vander}),
while the high T one is almost coinciding with the SAW one:
($(\gamma -\gamma_{1})_{SAW}=25/64$\cite{carlo}).

In $3D$, for a homopolymer, $\nu_{\Theta}$ is expected to
be equal to $1/2$ with logarithmic corrections \cite{sal}. 
Indeed $d=3$ is the upper critical dimension for the transition. 
We applied our methods to our model of random polyampholytes
in $3D$ and computed exact averages for chains up to 15 monomers.
A simple analysis of the radius of gyration,
not including logarithmic corrections, gives 
$\nu_{\Theta}=0.51 \pm 0.04$, again consistent with the
homopolymer universality class.

Alltogether the above results give very strong
evidence that the collapse transition of the globally
neutral random polyampholyte model falls in the same universality
class as the $\Theta$-point of homopolymers.
Support to such a conclusion comes from the 
exponent determinations we were able to perform.
Further evidence comes
from our comparative analysis of entropic properties
in the case of quenched and annealed disorders. Our study of 
$\gamma - \gamma_1$ shows that annealed and quenched partition
functions start to deviate appreciably at some temperature
falling  definitively below the estimated $T_{\Theta}$.
In order to obtain a collapse with exponents different from
those of homopolymer models one should have conditions
such that the effect of quenched disorder become important above,
or, at least, at the collapse transition temperature.
The identification and investigation of models where such conditions could
possibly be realized remains an important open
issue in the field, whose solution would sensibly increment
our understanding of the possible role played by chain disorder in polymer
statistics.

With the exact enumeration methods developed for the study
of the $\Theta$-transition we could also perform an analysis
of how the actual partition function at fixed $\{q\}$, $Z_{\{q\}}$,  
deviates from its (annealed) average at low temperature.
Histograms of quantities like $Z_{\{q\}}(N,T)/Z_{(a)}(N,T)$
show very clearly a lack of self-averaging at T sufficiently lower 
than  $T_{\Theta}$. While in a range of high T including
$T_{\Theta}$ they are narrow peaked  around the 
value 1, for lower temperature they are quite broad.
At very low T such histograms  acquire a sparse structure and allow to
investigate folding properties of the model.

While for T approaching zero homopolymers
collapse to many compact conformations
with the same ground state energy, most heteropolymer
sequences usually collapse to very few lowest-energy conformations
(see e.g. \cite{ground}).
In general, in order to
well represent properties of real 
proteins, a  heteropolymer
model is expected to  
admit a unique compact
conformation with lowest-energy, i.e. a non-degenerate
ground state, at least for some sequences.
The percentage of sequences admitting a unique
ground-state for the H-P heteropolymer model is believed 
from numerical analysis to remain almost
constant as $N$ increases \cite{dill1}.
The H-P model is a SAW in which each monomer can have either a 
hydrophobic or a polar character, with short range
interactions to the nearest neighbor solvent molecules.
This model has been often applied to protein folding studies
(see e.g. \cite{ground,dill2}).
Here  we investigated the number $f_N$ 
of sequences having a unique ``native state'' in our
$2D$ model.
This analysis was performed by applying the exact method described above
to the investigation of ground states of Hamiltonian walks
\cite{shak} on the square lattice, for chain lengths up to 25.
It turns out that 
$f_N$ grows with $N$ at 
a reduced exponential
rate with respect to the total number of sequences $N_q$.
In particular we found: $f_N\simeq 1.85^N$ while $N_q \simeq N^{-1/2} 2^N$.
Thus, the percentage of sequences which possess a unique ground state
tends asymptotically  to zero as $N \to \infty$.
This behavior is in sharp contrast with 
that found in the H-P model\cite{dill1}.

We thank F. Seno for useful discussions and help in the analysis.
We are also indebted with S. G. Whittington for discussions
and with C. Vanderzande for a critical reading of the manuscript.
Partial support from the European Network Contract 
ERBFMRXCT 980183 is also acknowledged.

\newpage

\centerline{\large FIGURE CAPTIONS}

\vskip 0.4 truein
\noindent
Fig 1:
Effective $\nu(N,K=2,T)$ exponents (solid lines)
and their linear extrapolation $\nu_{\infty}(T)$ 
for $1/N \to 0$ (dot-dashed line).
Temperature is normalized to monomer--monomer
interaction.
For sufficiently low T the sequences cease to be monotonic. 
Of course, the relatively short length of the chains rounds off
the expected step-like shape of $\nu_{\infty}$ at
the $\Theta$--transition. 
\vskip 0.4 truein
\noindent
Fig 2: 
Values of $\nu_{int}$ as a function of $1/N_{eff}$.
Rhombs indicate the means of the exponent extimates at fixed $N_{eff}$ 
, while horizontal bars limit the variance of their distribution.
\vskip 0.4 truein
\noindent
Fig 3:
Extrapolation of the crossover exponent $\phi_{\Theta}$.
\vskip 0.4 truein
\noindent
Fig 4:
Extrapolation of $\gamma-\gamma_1$ for annealed 
(dashed line) and quenched (dot-dashed line) 
disorder. The values are almost identical in a range of temperatures
extending below $T_{\Theta}\sim 0.80$.
\vskip 0.4 truein
\noindent
Tab 1:
Comparison between the number of different contact maps $S_N$ and
$C_N$ in $2D$ for $N=7,8,..,21$. Also even values of $N$ are reported for
completeness. 
\noindent
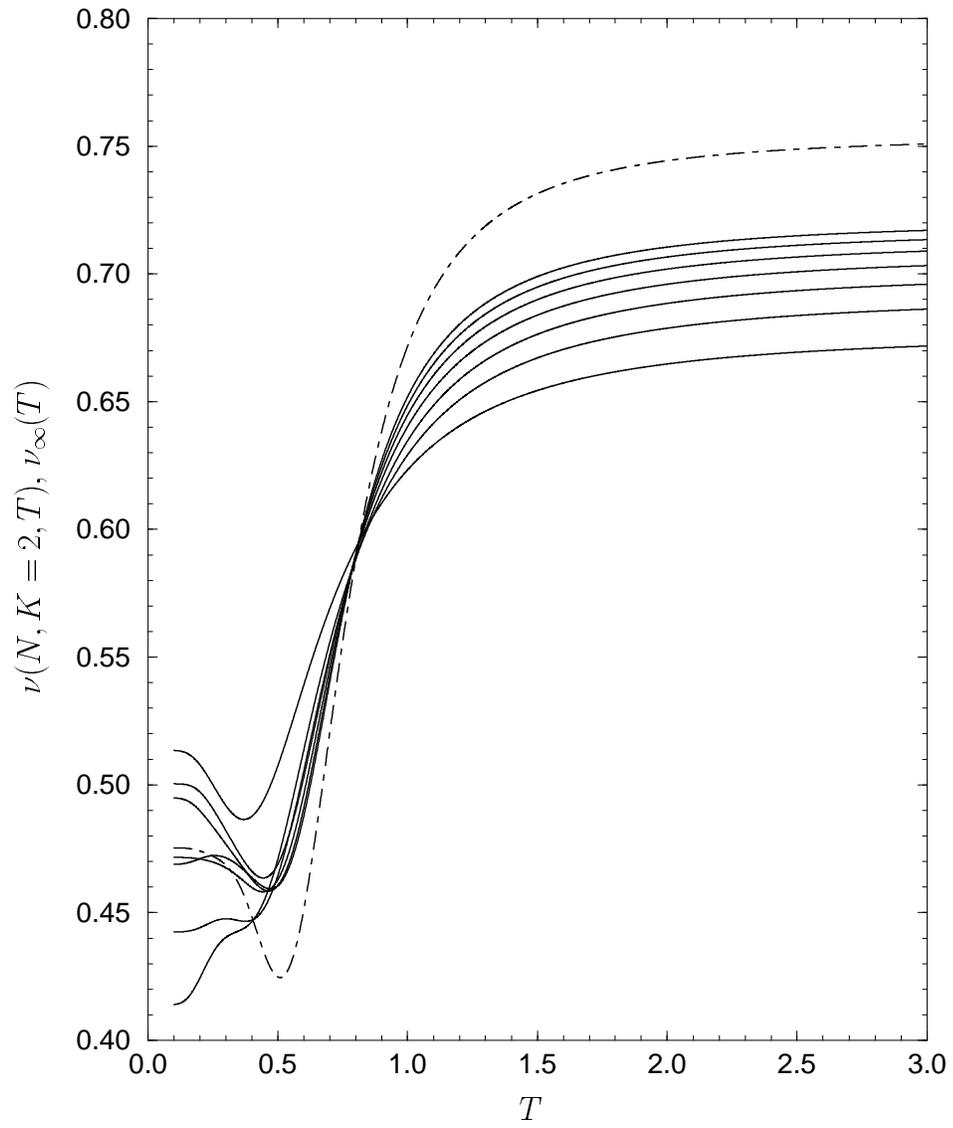
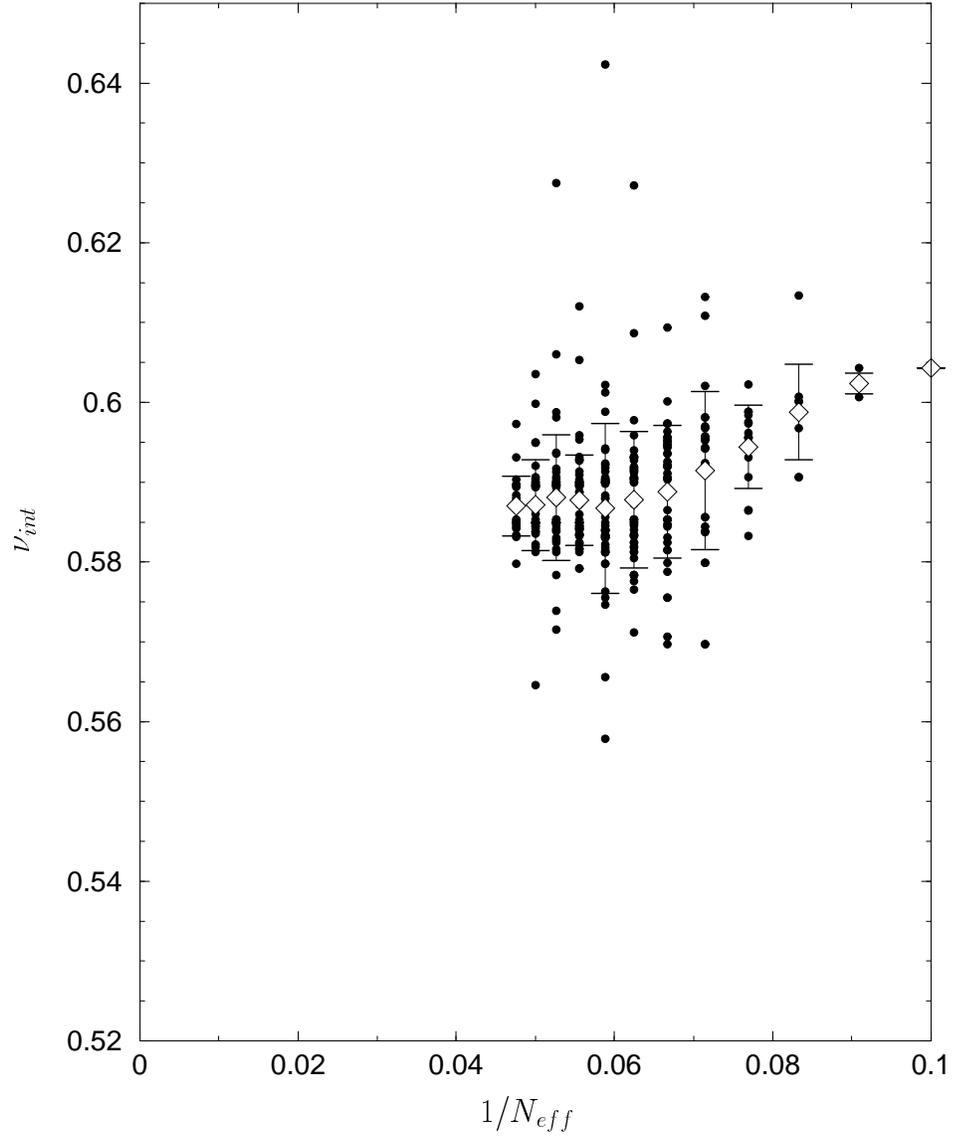
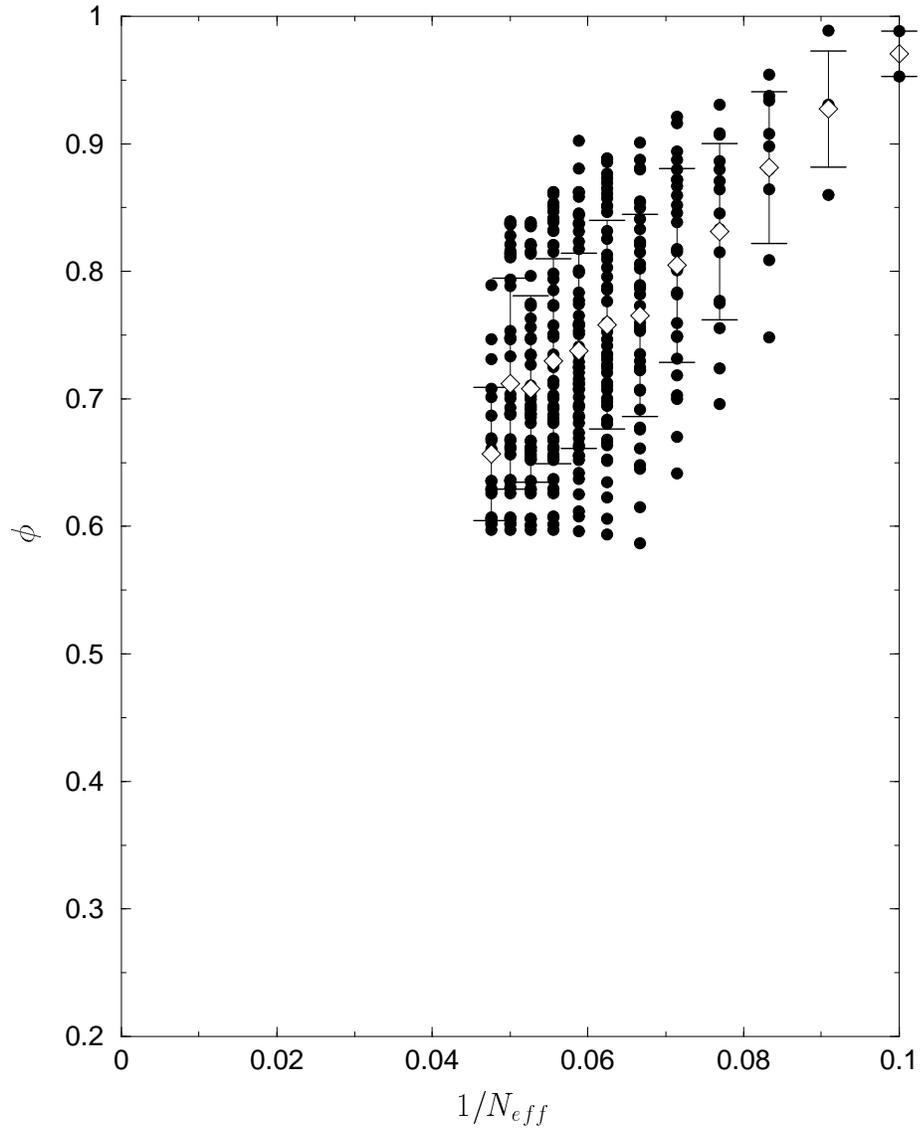
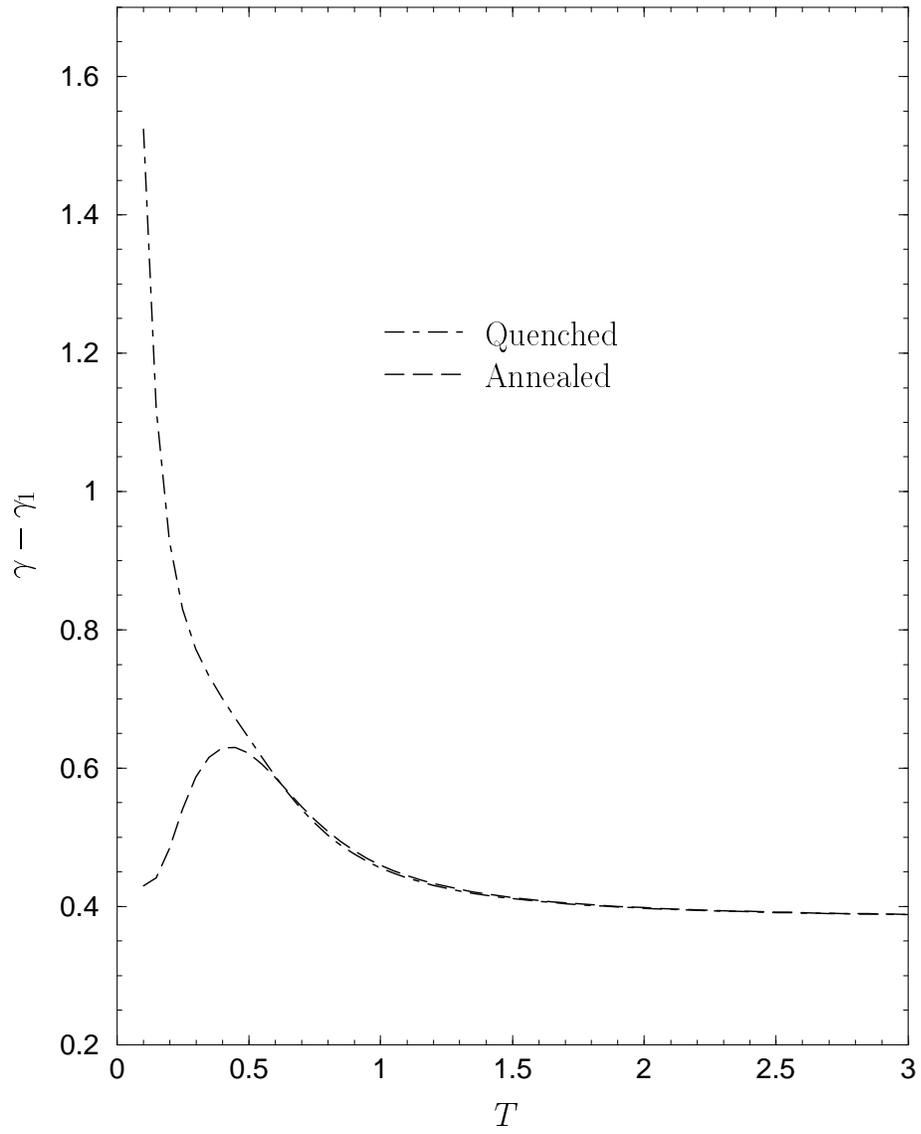
\end{document}